\documentstyle[12pt]{article}
\newcommand{\be}{\begin{equation}}
\newcommand{\ee}{\end{equation}}
\newcommand{\ba}{\begin{eqnarray}}
\newcommand{\ea}{\end{eqnarray}}

\begin{document}
\begin{center}
{\bf TWO-DIMENSIONAL SCHR\"ODINGER HAMILTONIANS WITH EFFECTIVE MASS IN SUSY APPROACH}\\
\vspace{0.5cm} {\large \bf F. Cannata$^{1,}$\footnote{E-mail:
cannata@bo.infn.it}, M.V. Ioffe$^{2,}$\footnote{E-mail:
m.ioffe@pobox.spbu.ru} ,
D.N. Nishnianidze$^{2,3,}$\footnote{E-mail: qutaisi@hotmail.com}}\\
\vspace{0.2cm}
$^1$ INFN, Via Irnerio 46, 40126 Bologna, Italy.\\
$^2$ Department of Theoretical Physics, Sankt-Petersburg State University,\\
198504 Sankt-Petersburg, Russia\\
$^3$ Akaki Tsereteli State University, 4600 Kutaisi, Georgia
\end{center}
\vspace{0.2cm} \hspace*{0.5in}
\hspace*{0.5in}
\begin{minipage}{5.0in}
{\small The general solution of SUSY intertwining relations of first order for two-dimensional Schr\"odinger
operators with position-dependent (effective) mass is built in terms of four arbitrary functions.
The procedure of separation of variables for the
constructed potentials is demonstrated in general form. The generalization for intertwining of second order is also
considered. The general solution for a particular form of intertwining operator is found, its properties - symmetry, irreducibility, separation of variables - are investigated.
\\
\vspace*{0.1cm} PACS numbers: 03.65.-w }
\end{minipage}
\vspace*{0.2cm}
\section{\bf Introduction.}
\vspace*{0.1cm} \hspace*{3ex} The non-relativistic quantum models with position-dependent mass were used
many years ago for the effective description of different models which are more complicated than the standard
"academic" one-particle Schr\"odinger equation (as examples, see \cite{old}). In this sense, position-dependent
mass is frequently dubbed as effective mass (EM). During the last years this sort of models became again very
fashionable in the literature, mainly due to the growing interest in such complex problems as
theoretical descriptions of nanodevices, motion in curved spaces, and models with
pseudo-Hermitian Hamiltonians (for illustration, see \cite{fashion}).

Different approaches to this branch of modern Quantum Mechanics were explored, and supersymmetric (SUSY) ones are among
the most promising. In the framework of SUSY method, one may try to find functional dependence
of effective mass and potential such that the corresponding Schr\"odinger equation is solvable or
quasi-exactly (partially) solvable. These solvable models may be considered both as a base for perturbation
expansions and as a laboratory for the study of qualitative properties. The SUSY method itself
in Quantum Mechanics includes many different variants \cite{susy}, and the SUSY intertwining relations seem
to be the most promising among them.

In most general form, SUSY intertwining relations between a pair of partner Hamiltonians $H_{1,2}$ are:
\ba
H_1 Q^+=Q^+H_2; \label{intertw1}\\
Q^-H_1=H_2Q^-, \label{intertw2}
\ea
where the mutually Hermitian conjugate intertwining operators $Q^{\pm}$ are called supercharges. These relations
lead to the isospectrality of Hamiltonians $H_{1,2}$ up to possible zero modes of supercharges. This means that (again, up to zero modes of $Q^{\pm}$) the energy spectra of $H_1$ and $H_2$ coincide.
Their bound state eigenfunctions are related (up to normalization factors) by the supercharges:
\be
H_i\Psi^{(i)}_n(\vec x)=E_n\Psi_n^{(i)}(\vec x);\quad i=1,2;\quad n=0,1,2,...;\quad
\Psi_n^{(2)}=Q^-\Psi_n^{(1)};\quad \Psi_n^{(1)}=Q^+\Psi_n^{(2)}.
\label{schr}
\ee
If either $Q^+$ or $Q^-$ have some zero modes, and they coincide with the wave functions
either of $H_2$ or of $H_1,$ these wave functions are annihilated according to (\ref{schr})
and have no analogous states in the spectra of the partner Hamiltonian.
Thus, from the intertwining relations one can find a pair of almost
isospectral Hamiltonians, which sometimes can be solved (partially or exactly).
By solving for the Hamiltonian we mean finding analytically its energy spectrum and corresponding wave functions.

We stress that the intertwining relations approach described above is very general, providing connections between pairs of spectral problems $H_1\, , H_2$.
It depends neither on specific nature of spectral problems of the operators $H_1,\, H_2,$ nor on
the specific form of intertwining operators $Q^{\pm}$ \cite{second}, \cite{third}, \cite{david}, \cite{junker}.

Each Hamiltonian $H_i$ included in the intertwining relations (\ref{schr}) has at least one symmetry operator $R_1=Q^+Q^-,\, R_2=Q^-Q^+,$
which commutes with it:
\be
[H_i, R_i]=0;\quad i=1,2
\label{symm}
\ee
Sometimes, these symmetry operators $R_i$ are expressed in terms of the Hamiltonian $H_i$ itself, and thus give no new information about the system. But otherwise, they describe indeed the symmetry of the model \cite{david}, \cite{junker}.

A variety of different realizations of the intertwining relations method was built during the development of SUSY
Quantum Mechanics. In the original one-dimensional SUSY Quantum Mechanics with supercharges given by first order differential operators, not more than one normalizable zero mode of $Q^{\pm}$ may exist,
and the spectra of $H_1,\, H_2$ either coincide or differ by one bound state. The "symmetry operators"
$R_i$ coincide with the Hamiltonians in this case due to their factorization property: $H_1=Q^+Q^-;\,
H_2=Q^-Q^+.$

In multi-dimensional SUSY Quantum Mechanics developed in \cite{abei} the supercharges of first order in derivatives
are used also, but a set of $d$ related intertwining relations between two scalar and $(d-1)$ matrix Hamiltonians
($d$ is dimensionality of space) must be considered simultaneously\footnote{Alternatively, intertwining of first order
between multi-dimensional {\it scalar} Hamiltonians leads \cite{kuru} to the systems amenable to separation of variables.}.
A more interesting realization of intertwining
relations was
constructed for one- and two-dimensional systems by using supercharges of
second \cite{second}, \cite{david}, \cite{junker} (and higher \cite{third}) orders in derivatives (Higher
order or Nonlinear SUSY Quantum Mechanics). In particular, for these systems the number of zero modes of $Q^{\pm}$
may be more than one (their number and properties are under the control). In the case of $d=2,$ many integrable
systems with nontrivial symmetry operators $R_i$ were built \cite{david}. Most of these systems are
not amenable to separation of variables.

The models with effective mass were also incorporated in SUSY intertwining relations approach
\cite{one}, but until now, all these papers concerned one-dimensional problems.
To our knowledge, the only work on higher-dimensional case was the paper of C.Quesne \cite{quesne}.
In this paper, the general system of equations equivalent to SUSY intertwining with scalar supercharges of first order in derivatives
was illustrated by a particular two-dimensional model, where the effective mass depends on one of Cartesian
coordinates only: $M(\vec x)=M(x_1).$ In this case, the EM Schr\"odinger equation certainly allows the separation of variables
in Cartesian coordinates, and the initial problem reduces to a pair of one-dimensional Schr\"odinger equations. It would be interesting to extend this approach to arbitrary form of EM function $M(\vec x)$ and to higher orders of supercharges.

The organization of the present paper is the following. In Section 2 the general intertwining relations of first order for
two-dimensional EM Hamiltonians are fully solved in terms of four arbitrary functions.
The mass functions, potentials and coefficients in the intertwining operators (supercharges) are built
explicitly in terms of these arbitrary functions.
We show that for the most general solution both intertwined Schr\"odinger Hamiltonians allow separation
of variables in suitable coordinates $y_1,y_2.$ Section 3 concerns SUSY
intertwining of second order. After general formulas are derived,
several models with specific forms of supercharges are analyzed. The general solution for a particular structure of supercharge (the case of elliptic metric) is obtained. In Conclusions the separation of variables as well as the symmetry properties of the system (its integrability) and irreducibility of supercharge are discussed.

\section{\bf General solution of first order intertwining relations.}
\vspace*{0.1cm} \hspace*{3ex}
The two-dimensional Schr\"odinger operator with effective mass $m(\vec x)\equiv m_0M(\vec x)$
and $\hbar = 2m_0=1$ can be written in explicitly Hermitian form \cite{roos} as\footnote{Everywhere
below the summation over repeated indices is implied.}
\begin{equation}\label{roos}
  H=-\frac{1}{2}[M^{\alpha}(\vec x)\partial_iM^{\beta}(\vec x)\partial_iM^{\gamma}(\vec x) +
  M^{\gamma}(\vec x)\partial_iM^{\beta}(\vec x)\partial_iM^{\alpha}(\vec x)] + V(\vec x),
\end{equation}
where $\vec x\equiv (x_1,x_2),\,\,\partial_i\equiv \partial /\partial x_i,$ the effective
mass function $M(\vec x)$ is dimensionless function of coordinates, the real function $V(\vec x)$ is the potential, and
$\alpha , \beta , \gamma$ are constant parameters
satisfying the physically motivated restriction $\alpha + \beta +\gamma = -1.$
The specificparticular values for $\alpha , \beta , \gamma$ depend on the specific physical
model which is described by the EM Schr\"odinger equation.

It is convenient to rewrite (\ref{roos}) in the form:
\be
H=-\partial_i\frac{1}{M(\vec x)}\partial_i + U(\vec x),
\label{eff}
\ee
which unifies different choices of $\alpha , \beta , \gamma$ in the additional terms of effective potential
$U(\vec x):$
\begin{equation}
U(\vec x)= V(\vec x)+\frac{\beta +1}{2}\frac{\Delta^{(2)}M(\vec x)}{M^2(\vec x)}-
[\alpha(\alpha + \beta +1)+\beta +1]\frac{(\partial_iM(\vec x))(\partial_iM(\vec x))}{M^3(\vec x)};\quad
\Delta^{(2)}\equiv \partial_i\partial_i.
\label{UV}
\end{equation}

Let us suppose that a pair of partner two-dimensional Hamiltonians $H_1,\, H_2$ of the form (\ref{eff}),
both with effective mass $M(\vec x),$ and potentials
$U_1(\vec x)\, , U_2(\vec x)$,  satisfy the intertwining relations
(\ref{intertw1}), (\ref{intertw2}) with most general first order supercharges:
\begin{equation}\label{Q+}
  Q^+=(Q^-)^{\dagger}=q_l(\vec x)\partial_l+q(\vec x); \quad l=1,2 .
\end{equation}
The explicit form of the intertwining relations (\ref{intertw1}), (\ref{intertw2}) seems to be too
complicated in the two-dimensional case to be solved analytically. Only choosing some ansatzes for
$M(\vec x)$ and $V(\vec x)$ one can simplify the problem to make it solvable: this was done in paper \cite{quesne}.

In order to find the {\bf general} solution of intertwining relations one must make their form
more tractable.
First of all, it is useful to eliminate the first derivatives from (\ref{eff}) a suitable similarity transformation:
\ba
h_{1,2}&=&e^{-\phi (\vec x)}H_{1,2} e^{\phi (\vec x)}\equiv -\frac{1}{M(\vec x)}\Delta^{(2)}+v_{1,2}(\vec x);
\label{similarity}\\
v_{1,2}(\vec x)&=&U_{1,2}(\vec x)-\frac{\Delta^{(2)}M(\vec x)}{2M^2(\vec x)}+
\frac{3\biggl( \partial_kM(\vec x) \biggr)\biggl( \partial_kM(\vec x)\biggr)}{2M^3(\vec x)} \label{vU}
\ea
with some real function $\phi (\vec x),$ and
\begin{equation}\label{similarity2}
  q^{\pm}=e^{-\phi (\vec x)}Q^{\pm}e^{\phi (\vec x)}=q_l(\vec x)\partial_l+p(\vec x);\quad
p(\vec x)=q(\vec x)+\frac{q_k(\vec x)\biggl( \partial_kM(\vec x) \biggr)}{2M(\vec x)}.
\end{equation}
The function $\phi (\vec x)$ can be determined from the condition of absence of first derivatives in $h_{1,2}:$
\be
\partial_k\phi (\vec x)=\frac{\biggl( \partial_kM(\vec x)\biggr)}{2M(\vec x)};\quad
\phi (\vec x)=\frac{1}{2}\ln M(\vec x) + Const.
\label{phi}
\ee
This similarity transformation certainly is not unitary, and therefore,
$h_{1,2}$ are not Hermitian, and $q^+\neq (q^-)^{\dagger}.$ Nevertheless, these
operators satisfy the intertwining relation equivalent to (\ref{intertw1}):
\be
h_1 q^+=q^+h_2.
\label{intertw}
\ee
Equivalence means that solutions of (\ref{intertw1}) and of (\ref{intertw}) are in one-to-one
correspondence. Furthemore, solution of (\ref{intertw2}) will be automatically obtained, being the Hermitian conjugate of (\ref{intertw1}).

Thus, we will study relation (\ref{intertw}), whose solution seems to be technically much simpler than (\ref{intertw1}).
It can be rewritten as a system of six nonlinear differential equations for the functions $M,\, v_{1,2},\,q_i,\, p:$
\ba
&&M(\vec x)\biggl[ \biggl( \partial_kq_i(\vec x)\biggr)+\biggl(\partial_iq_k(\vec x) \biggr) \biggr] +
\delta_{ik}q_j(\vec x)\biggl( \partial_jM(\vec x) \biggr)=0;
\label{1}\\
&&2M(\vec x)v(\vec x)q_i(\vec x)-2\biggl( \partial_ip(\vec x)\biggr)-\biggl( \Delta^{(2)}q_i(\vec x)\biggr)=0;
\label{2}\\
&&M(\vec x)\biggl[ q_i(\vec x)\biggl( \partial_iv_2(\vec x) \biggr)-2v(\vec x)p(\vec x)\biggr]+
\biggl( \Delta^{(2)}p(\vec x)\biggr)=0,
\label{3}
\ea
where all space indices take two values $1,2$, and the function $v(\vec x)$ is defined as:
\be
2v(\vec x)\equiv v_1(\vec x)-v_2(\vec x). \label{vvv}
\ee

The general solution of the three Eqs.(\ref{1}) can be derived as follows. After subtraction of equations for $i=k,$ one obtains that $\partial_1q_1=\partial_2q_2,$ i.e. $q_1=\partial_1F(\vec x);\,
q_2=\partial_2F(\vec x).$ Then
equation for $i\neq k$ gives that the real function $F(\vec x)$ satisfies $\Delta^{(2)}F(\vec x)=0.$
Therefore, the coefficients $q_1, q_2$ are combinations of analytical functions $f(z)$ and $f^{\star}(z^{\star})$
of $z=x_1+ix_2$ and
$z^{\star} =x_1-ix_2,$ correspondingly:
\be
q_1(\vec x)=i(f(z)- f^{\star}(z^{\star}));\quad q_2(\vec x)=f(z)+ f^{\star}(z^{\star}). \label{q}
\ee
After that, the last equation of (\ref{1}) takes the form:
\be
(f\partial_z-f^{\star}\partial_{z^{\star}})M(\vec x)=-(f^{\prime}- f^{\star\prime})M(\vec x).
\ee
Its general solution can be found by the substitution $M^{-1}(\vec x)
\equiv \frac{1}{4}(f(z) f^{\star}(z^{\star}))\Omega (\vec x),$ because $\Omega (\vec x) $
must satisfy the first order equation:
\be
(f\partial_z-f^{\star}\partial_{z^{\star}})\Omega (\vec x)=0.
\ee
Thus, the general solution of (\ref{1}) consists of Eqs.(\ref{q}) and:
\be
M^{-1}(\vec x)=\frac{1}{4}f(z)f^{\star}(z^{\star})\Omega \biggl(\int \frac{dz}{f(z)}+
\int \frac{dz^{\star}}{f^{\star}(z^{\star})}\biggr),
\label{Omega}
\ee
where $\Omega $ is an arbitrary real function of its argument.

The last terms in both equations in (\ref{2}) disappear due to (\ref{q}). Then Eq.(\ref{2}) takes the form:
\be
\biggl( \partial_{z^{\star}}p(\vec x)\biggr)= iM(\vec x)v(\vec x)f(z);\quad
\biggl( \partial_{z}p(\vec x)\biggr) = -iM(\vec x)v(\vec x) f^{\star}(z^{\star}).
\label{qq}
\ee
Therefore, $p(\vec x)$ is expressed in terms of an arbitrary real function $P$ of one specific real variable:
\be
p(\vec x)=4P\biggl(i(\int \frac{dz}{f(z)}-\int \frac{dz^{\star}}{f^{\star}(z^{\star})})\biggr),
\label{p}
\ee
and the function $v(\vec x)$ has a compact form in terms of $\Omega$ and $P:$
\begin{equation}\label{vP}
  2v(\vec x)=v_1-v_2=-2P^{\prime}\biggl(i(\int \frac{dz}{f(z)}-\int \frac{dz^{\star}}{f^{\star}(z^{\star})})\biggr)
  \Omega \biggl(\int \frac{dz}{f(z)}+
\int \frac{dz^{\star}}{f^{\star}(z^{\star})}\biggr),
\end{equation}
where $P^{\prime}$ means the derivative in respect to its argument.

It is evident now that, instead of the Cartesian arguments $x_{1}, x_{2},$ the alternative real space coordinates $y_1, y_2$
are much more convenient:
\begin{equation}\label{y}
y_1\equiv \int \frac{dz}{f(z)}+\int \frac{dz^{\star}}{f^{\star}(z^{\star})};\quad
y_2\equiv i\biggl(\int \frac{dz}{f(z)}-\int \frac{dz^{\star}}{f^{\star}(z^{\star})}\biggr).
\end{equation}

Finally, taking into account the previous results, the remaining equation (\ref{3}) admits the general solution:
\begin{equation}\label{v2}
  v_2(\vec y)=\Omega (y_1)  \biggl(P^2(y_2)+   P^{\prime}(y_2)\biggr)  + \Gamma (y_1),
\end{equation}
where a new arbitrary real function $\Gamma $ is introduced.

Summarizing the obtained results: we derived the general solution of intertwining relations (\ref{intertw}),
represented in (\ref{q}), (\ref{Omega}), (\ref{p}), (\ref{vP}) and (\ref{v2}) in terms of
an arbitrary analytical function $f(z)$ and three arbitrary real functions $\Omega (y_1), P(y_2), \Gamma (y_1),$
restricted by physical reasons only: for example, $\Omega$ must be positive.
The use of variables $y_1, y_2$ allows to rewrite the operators $h_{1,2}$ from (\ref{similarity}) in a rather compact form:
\begin{equation}\label{hy}
  h_{1,2}= -\frac{1}{4}\Omega(y_1)\Delta_y^{(2)}+\Omega(y_1)\biggl(P^2(y_2)\mp P^{\prime}(y_2)\biggr)+\Gamma (y_1);\quad \Delta_y^{(2)}\equiv \partial_{y_i}\partial_{y_i}.
\end{equation}
The intertwining operator $q^+$ from (\ref{q}), (\ref{p}) is also simplified essentially by use of $y_1, y_2:$
\begin{equation}\label{qqq}
  q^+=4(-\partial_{y_2}+P(y_2)),
\end{equation}
i.e. it depends on $y_2$ only, having a form typical for one-dimensional SUSY Quantum Mechanics with constant mass.
The operators $h_{1,2}$ with "mass" $4\Omega^{-1}(y_1)$ are intertwined (only in the variable $y_2$) by the operators
$q^+,$ therefore their eigenvalues and eigenfunctions are
interrelated. Performing the inverse similarity transformation to the physical Hamiltonians $H_{1,2},$ one can
link their spectral properties by $Q^{\pm}$ as well.

The variables $y_1, y_2$ are useful not only to express the most general form of intertwined Hamiltonians $H_{1,2},$
but also to formulate the algorithm of explicit separation of variables in the spectral equations for
operators $h_{1,2}$ (analogues of physical stationary Schr\"odinger equations for Hamiltonians $H_{1,2}$):
\begin{equation}\label{sschr}
\biggl[-\frac{1}{4}\Omega(y_1)\Delta_y^{(2)}+
\Omega(y_1)\biggl(P^2(y_2)\mp P^{\prime}(y_2)\biggr)+\Gamma (y_1)\biggr]\psi^{(1,2)}_n=E_n\psi^{(1,2)}_n.
\end{equation}
Indeed, dividing (\ref{sschr}) by $\frac{1}{4}\Omega(y_1)$ one may search for its solutions,
skipping for simplicity upper indices $1,2,$ as products
\begin{equation}\label{product}
\psi_n=\eta_n(y_1)\rho_n(y_2)
\end{equation}
of solutions of two one-dimensional problems:
\ba
&&\biggl[-\frac{1}{4}\partial_{y_2}^2+\biggl(P^2(y_2)\mp P^{\prime}(y_2)\biggr)\biggr]\rho_n(y_2)=\epsilon_n\rho_n(y_2);
\label{second}\\
&&\biggl[-\frac{1}{4}\partial_{y_1}^2+
\biggl( \Gamma (y_1)-E_n \biggr)\Omega^{-1}(y_1)\biggr]\eta_n(y_1)=-\epsilon_n\eta_n(y_1).
\label{first}
\ea
Notice that the "potential term" in (\ref{first}) depends explicitly on the "spectral parameter" of (\ref{second}).
Selecting the solutions $\eta_n(y_1), \rho_n(y_2)$ of (\ref{first}), (\ref{second}), one has to
remember that the conditions of normalizability may be changed by similarity transformation\footnote{It is clear from (\ref{phi}) that normalizability is not changed if the mass function $M(\vec x)$ is everywhere positive finite function.}:
\begin{equation}\label{Psi}
\Psi_n=e^{\phi}\psi_n= e^{\phi}\eta_n(y_1)\rho_n(y_2).
\end{equation}
Thus, solving two-dimensional Schr\"odinger equations with Hamiltonians $H_{1,2}$ by separation of variables for $h_{1,2}$,
one has to find discrete values $E_n,$ such that the functions (\ref{Psi}), in turn,
constructed  from solutions of two
one-dimensional problems (\ref{first}), (\ref{second}), are normalizable.

It is clear that the construction above starts from a choice of the analytical function $f(z).$ In particular,
it is instructive to consider the simplest case $f(z)=f^{\star}(z^{\star})= Const ,$ where,
because of (\ref{y}),
$y_1=x_1, \, y_2=x_2$ up to a change of scale. In the same time, the mass $M(\vec x)$ depends on $x_1$ only, due to
the general solution (\ref{Omega}). Thus, in this case the two-dimensional operators $h_{1,2}$
in (\ref{intertw}) are intertwined by $q^+$ only in one variable $x_2,$ on which the mass does not depend
$M(\vec x)=M(x_1).$
The two-dimensional problem (\ref{intertw1}) becomes fully one-dimensional. The analogous result
(up to replacing $x_1\leftrightarrow x_2$) would be obtained for the choice $f(z)=-f^{\star}(z^{\star})=Const,$
where $M(\vec x)=M(x_2).$

Just the models with mass $M(\vec x),$ depending only on one variable $x_1,$ were considered in paper \cite{quesne}. It is possible to derive the most general condition for functions
$f, f^{\star}, \Omega$ above, such that the mass $M(\vec x)$ does not depend on $x_2.$ From Eq.(\ref{Omega}), the equation $\biggl( \partial_2M(\vec x)\biggr)=0$ gives:
\be
\biggl( \ln \Omega(y_1) \biggr)^{\prime}=
\frac{f^{\prime}(z)f^{\star}(z^{\star})-f(z)f^{\star\prime}(z^{\star})}{f(z)-f^{\star}(z^{\star})}.
\label{ln}
\ee
The r.h.s. has to be independent on $y_2$ as well, leading to the functional-differential equation:
\be
f^{\prime\prime}f-f^{\star}f^{\prime\prime}-(f^{\prime})^2 -f^{\star\prime\prime}f^{\star}+
ff^{\star\prime\prime}+(f^{\star\prime})^2=0
\ee
with two possible solutions only:
\ba
&&f(z)=az^2+bz+c;
\label{polinom}\\
&&f(z)=\alpha e^{\lambda z}+\beta e^{-\lambda z}+\gamma.
\label{exp}
\ea
Depending on the choice of constants, many different functions $M(x_1)$ are possible. As example:
\be
f(z)=az^2;\quad y_{1,2}=- \frac{2x_{1,2}}{azz^{\star}};\quad \Omega(y_1)=\alpha y_1^2;\quad M(\vec x)=\frac{1}{\alpha x_1^2}.
\ee
Analogously (although more cumbersome), the choice $f(z)=(az+b)^2$ leads to one of the models
of \cite{quesne} with $M(\vec x)=\gamma/(ax_1+b)^2.$

One might be interested in models with variables $y_1, y_2,$ coinciding with polar
coordinates $\rho,\,\theta .$ This condition leads to
$f(z)=bz,$
with $b$ - an arbitrary real (or pure imaginary) constant. There is no need to satisfy (\ref{ln}) in this case:
$\Omega$ remains an arbitrary function of $y_1$, defining the form (\ref{Omega}) of mass function. In particular, for real value
of $b$ the mass $M(\vec x)$ is a function of $\rho$ only:
$M^{-1}(\vec x)=\frac{b^2}{4} \rho^2\Omega(\frac{2}{b} \ln\rho).$
In full analogy to the case of Cartesian coordinates above, the most general condition, when mass depends only on one of
polar coordinate, is wider than the simplest form $f(z)=bz.$

\section{\bf Second order intertwining.}
\vspace*{0.1cm} \hspace*{3ex}
As it was shown in the previous Section, the first order intertwining of the two-dimensional operators $h_{1,2}$
leads actually to separation of variables in terms of $y_1, y_2.$ And furthermore, after separation
one obtains the intertwining problem (\ref{intertw}) with a constant mass. This rather trivial result is not very surprising
in the light of the study of two-dimensional SUSY Quantum Mechanics (with constant mass), where
nontrivial achievements were obtained following two directions.
The first one \cite{abei} dealt with two-component (vector) first order supercharge operators, which intertwined a given
scalar Hamiltonian with matrix partner.
Alternatively, if one doesn't wish to deal with matrix potentials, a second option was
proposed \cite{david} - intertwining by scalar supercharges $q^+$ of second order in derivatives. This
latter approach leads us to use the second order operators in the problem with effective mass considered in the present paper.

The most general intertwining operator of second order in (\ref{intertw}) is:
\begin{equation}\label{q2}
  q^+=g_{ij}(\vec x)\partial_i\partial_j + C_i(\vec x)\partial_i + B(\vec x),
\end{equation}
where $g_{ij}, C_i, B$ are arbitrary real functions of $\vec x.$ Equating coefficients of each power in the derivatives
in (\ref{intertw}), one obtains a very complicated system of nonlinear differential equations for functions
$M(\vec x), v_{1,2}(\vec x)$ and all coefficient functions in (\ref{q2}). In order to have a chance to solve this system,
we make a simplifying ansatz for the metric in the supercharge:
\begin{equation}\label{g}
  g_{ik}(\vec x)=\omega(\vec x)\delta_{ik}.
\end{equation}

From third order terms in (\ref{intertw}) it follows that the mass and metric are related:
\begin{equation}\label{Mg}
  M(\vec x)=\omega^{-1}(\vec x).
\end{equation}

The second order terms give a system of three equations:
\be
M(\vec x)\biggl( \partial_iC_k(\vec x)+\partial_kC_i(\vec x)\biggr)=\biggl[2M^2(\vec x)v(\vec x)\omega(\vec x)
-C_j(\vec x)\biggl( \partial_jM(\vec x)\biggr)\biggr]\delta_{ik},
\label{dd}
\ee
where $v(\vec x)$ was defined in (\ref{vvv}). Two of these equations can be solved in terms of
an arbitrary analytical function $g(z)$ similarly to (\ref{q}):
\begin{equation}\label{C}
  C_1(\vec x)=i(g(z)-g^{\star}(z^{\star}));\quad   C_2(\vec x)=g(z)+g^{\star}(z^{\star}).
\end{equation}
The solution of the remaining equation in (\ref{dd}):
\begin{equation}\label{postpone}
2M(\vec x)\biggl[ \biggl(\partial_1C_1(\vec x)\biggr)-v(\vec x)\biggr]+
C_i(\vec x)\biggl( \partial_iM(\vec x)\biggr)=0.
\end{equation}
we postpone for later.

The first order terms can be written as:
\begin{equation}\label{d}
\partial_i\biggl( B(\vec x)+v_2(\vec x)\biggr)=M(\vec x)v(\vec x)C_i(\vec x).
\end{equation}
Instead of the second equation in (\ref{d}), it is more convenient to solve the special combination
of both equations, obtained by multiplying by $C_i$ and subtracting:
\be
\biggl(C_2(\vec x)\partial_1-C_1(\vec x)\partial_2\biggr)\biggl(B(\vec x)+v_2(\vec x)\biggr)\equiv
4\partial_{t_1}\biggl(B(\vec x)+v_2(\vec x)\biggr)=0,
\label{comb}
\ee
where we used the variables $t_{1,2}$ in full analogy to $y_{1,2},$ defined in (\ref{y}), up to the replacement $f(z)\leftrightarrow g(z)$.  Therefore:
\be
B(\vec x) + v_2(\vec x)=4S(t_2),
\label{MM}
\ee
with arbitrary function $S(t_2).$ After that, we express $\partial_1$ in the first equation of (\ref{d})
in terms of $\partial_{t_1}, \partial_{t_2}$ and use (\ref{MM}), deriving the relation between mass $M(\vec x)$ and $v(\vec x):$
\be
M(\vec x)=-\frac{4S^{\prime}(t_2)}{g(z)g^{\star}(z^{\star})v(\vec x)}.
\label{ccomb}
\ee

Now we can solve the postponed Eq.(\ref{postpone}), rewriting it in terms of the variables $t_1, t_2$ and using (\ref{C}) and (\ref{ccomb}). After some transformations, it takes the form of the well known Bernoulli
equation for the function $v(\vec t)$:
\be
\partial_{t_2}v(\vec t)=\frac{1}{2}v^2(\vec t)+v(\vec t)\frac{S^{\prime\prime}(t_2)}{S^{\prime}(t_2)}.
\label{bern}
\ee
Its general solution \cite{polyanin} (section 1.1.5) includes an integration constant, which becomes an arbitrary function $A(t_1)$ of $t_1$ in our case:
\be
v(\vec t)=-\frac{2S^{\prime}(t_2)}{A(t_1)+S(t_2)}.
\label{bernsolution}
\ee

The term without derivative operators in (\ref{intertw}) reads:
\be
\Delta^{(2)}\biggl( B(\vec x)+ v_2(\vec x) \biggr)+M(\vec x)C_i(\vec x)\biggl( \partial_iv_2(\vec x)\biggr)
-2M(\vec x)v(\vec x)B(\vec x)=0,
\label{d0}
\ee
and it can be transformed to the first order differential equation for $v_2(t_1,t_2)$ with coefficients
which were defined above:
\be
(\partial_{t_2}v_2(\vec t)) + \frac{S^{\prime}(t_2)}{\biggl( A(t_1)+S(t_2) \biggr)}v_2(\vec t) - \frac{2\biggl( S^{\prime}(t_2) +S^2(t_2) \biggr)^{\prime}}{\biggl( A(t_1)+S^(t_2) \biggr)}=0.
\label{d0d0}
\ee
The general solution for $v_2$ and its partner $v_1$ have the form:
\be
v_{1,2}(\vec t)=\frac{2}{S(t_2)+A(t_1)}\biggl[ S^2(t_2)\mp S^{\prime}(t_2)+D(t_1) \biggr],
\label{vv2}
\ee
where $D(t_1)$ is a new arbitrary function. Finally, the operators $h_{1,2}$ take the form:
\be
h_{1,2}=-\frac{2}{S(t_2)+A(t_1)}\Delta^{(2)}_y + \frac{2}{S(t_2)+A(t_1)}\biggl[ S^2(t_2)\mp S^{\prime}(t_2)+D(t_1) \biggr], \label{last}
\ee
allowing the separation of variables $t_1$ and $t_2$ in the corresponding Schr\"odinger equations,
analogous to (\ref{sschr}) of previous Section. But in contrast to Eqs.(\ref{second}), (\ref{first}),
the spectral parameter $E_n$ is present now in both one-dimensional equations, leading to essential difficulties in solving the spectral problem.

\section{\bf Conclusions.}
\vspace*{0.1cm} \hspace*{3ex}
As we discussed in the Introduction, an arbitrary interwining relation between physical (Hermitian)
Hamiltonians $H_{1,2}$ leads to the Hermitian symmetry operators $R_1=Q^+Q^-,\, R_2=Q^-Q^+.$ In the context
of Section 2, these operators are obviously of second order in derivatives. Meanwhile, in Section 3, they are
initially of fourth order. Since the symmetry operator is defined up to a function of Hamiltonian itself,
one must study the possibility to reduce the order of the symmetry. Because from the very beginning we restricted
ourselves to real coefficient functions in the supercharges $Q^{\pm},$ the terms of odd order in derivatives,
being non-Hermitian, cannot appear in $R_i.$  As for the term of fourth order in derivatives, due to relation
(\ref{Mg}) between metric of supercharge and mass function in $H_i,$
it can be just expressed as $H_i^2$ plus some terms of even (second and zero) orders. This means that
for the metric (\ref{g}) in the supercharge, systems with Hamiltonians $H_i$ obey symmetries of second order in
derivatives only. This fact is not surprising in the light of separation of variables in Schr\"odinger
equations discussed in the very end of Section 3.

An interesting question concerns also the reducibility of the considered second order supercharges, i.e. the possibility in principle to factorize the operator $q^+$ onto two multipliers of first order of the form (\ref{similarity2}). It is useful to compare the terms of highest (second) order in (\ref{q2}), (\ref{g})
and in (\ref{similarity2}), (\ref{q}). After a simple analysis (in particular, due to absence of mixed term $\partial_1\partial_2$ in (\ref{q2})), it becomes clear that such factorization is possible only in the case when both the functions $f_1(z), f_2(z)$  in the first order multipliers and $\omega(\vec x)$ in (\ref{g}) are constants. This situation corresponds to the case of constant mass $M(\vec x)$ and is not relevant for the present discussion. Thus, the supercharges $q^+$ of the form
(\ref{q2}), (\ref{g}) are irreducible. The same conclusion can be reached after comparing the general expressions for the intertwined Hamiltonians (\ref{hy}) and (\ref{last})
for the first and second order intertwinings. The denominator $\biggl( S(t_2)+A(t_1) \biggr),$ present in both kinetic
and potential terms for second order intertwining, may appear in the expressions (\ref{hy}) for Hamiltonians,
involved in two consecutive first order intertwinings, for the case of constant mass only. This argumentation
corroborates the conclusion about irreducibility of second order $q^+$ from another point of view.

\section*{\bf Acknowledgments}
The work was partially supported by INFN (M.V.I. and D.N.N.) and by the Russian grants
RFFI 06-01-00186-a, RNP 2.1.1.1112 (M.V.I.).  

\vspace{.2cm}

\begin{thebibliography}{}
\bibitem{old}
G.H.Wannier, Phys. Rev. 52(1937) 191;\\
J.C.Slater, Phys. Rev. 76 (1949) 1592;\\
J.M.Luttinger, W.Kohn, Phys. Rev. 97 (1955) 869;\\
M.A.Preston, Physics of the Nucleus, Addison-Wesley, INC, Reading, 1965, p.210.
\bibitem{fashion}
C.Weisbuch, B.Vinter, Quantum Semiconductor Heterostructures, Academic Press, New York, 1997;\\
L.Serra, E.Lipparini, Eur. Phys. Lett. 40 (1997) 667;\\
L.Dekar, L.Chetouani, T.F.Hammann, Phys. Rev. A59 (1999) 107;\\
A.de S.Dutra, C.A.S.Almeida, Phys. Lett., A275 (2000) 25;\\
C.Quesne, V.M.Tkachuk, J. Phys. A37 (2004) 4267;\\
A.Ganguly, M.V.Ioffe, L.M.Nieto, J. Phys. A39 (2006) 14659;\\
C.M.Bender, Rep. Prog. Phys. 70 (2007) 947.
\bibitem{susy}
E.Witten, Nucl. Phys. B188 (1981) 513;\\
F.Cooper, A.Khare, U.Sukhatme, Phys.Rep. 251 (1995) 268;\\
G.Junker, Supersymmetrical Methods in Quantum and Statistical Physics, Springer, Berlin, 1996;\\
B.K.Bagchi, Supersymmetry in Quantum and Classical Mechanics, Chapman, Boca Raton, 2001.
\bibitem{second}
A.A.Andrianov, M.V.Ioffe, V.P.Spiridonov, Phys.Lett. A174 (1993) 273;\\
A.A.Andrianov, F.Cannata, J.-P.Dedonder, M.V.Ioffe, Int.J.Mod.Phys. A10 (1995) 2683.
\bibitem{third}
M.V.Ioffe, D.N.Nishnianidze, Phys.Lett. A327 (2004) 425.
\bibitem{david}
A.A.Andrianov, M.V.Ioffe, D.N.Nishnianidze, Phys.Lett. A201 (1995) 103;\\
A.A.Andrianov, M.V.Ioffe, D.N.Nishnianidze, Theor. and Math.Phys. 104 (1995) 1129;\\
F.Cannata, M.V.Ioffe, D.N.Nishnianidze, J.Phys. A35 (2002) 1389;\\
M.V.Ioffe, J.Phys. A37 (2004) 10363;\\
M.V.Ioffe, P.A.Valinevich, J.Phys. A38 (2005) 2497;\\
M.V.Ioffe, D.N.Nishnianidze, Phys. Rev. A76 (2007) 052114.
\bibitem{junker}
F.Cannata, M.V.Ioffe, G.Junker, D.N.Nishnianidze, J. Phys. A32 (1999) 3583.
\bibitem{abei}
A.A.Andrianov, N.V.Borisov, M.I.Eides, M.V.Ioffe, Phys. Lett. A109 (1985) 143.
\bibitem{kuru}
S.Kuru, A.Tegmen, A.Vercin, J. Math. Phys. 42 (2001) 3344;\\
B.Demircioglu, S.Kuru, M.Onder, A.Vercin, J. Math. Phys. 43 (2002) 2133.
\bibitem{one}
A.R.Plastino, A.Rigo, M.Casas, F.Garsias, A.Plastino, Phys. Rev. A60 (1999) 4318;\\
V.Milanovic, Z.Ikonic, J.Phys. A32 (1999) 7001;\\
B.Gonul, O.Ozer, B.Gonul, F.Uzgum, Mod. Phys. Lett. A17 (2002) 2453;\\
R.Koc, M.Koca, J. Phys. A36 (2003) 8105;\\
B.Gonul, M.Koak, Mod. Phys. Lett. A20 (2005) 355;\\
B.Bagchi, P.Gorain, C.Quesne, R.Roychoudhury, Eur. Phys. Lett. 72 (2005) 155;\\
A.Ganguly, L.M.Nieto, J. Phys. A40 (2007) 7265
\bibitem{quesne}
C.Quesne, Ann. Phys. 321 (2006) 1221
\bibitem{roos}
O.von Roos, Phys. Rev. B27 (1983) 7547.
\bibitem{polyanin}
A.D.Polyanin, V.F.Zaitsev, Handbook of Exact Solutions for Ordinary Differential Equations,
Chapmann and Hall/CRC, 2002.

\end{thebibliography}
\end{document}